\documentclass[twocolumn,showpacs,preprintnumbers,amsmath,amssymb]{revtex4}
\usepackage{graphicx} 
\usepackage{dcolumn}  
\usepackage{bm}       
\begin{document} 
\title{Theory of the Lightly Doped Mott insulator}
\author{R. Eder$^{1,2}$, P. Wr\'obel$^3$ and Y. Ohta$^2$}
\affiliation{$^1$Karlsruhe Institut of Technology,
Institut f\"ur Festk\"orperphysik, 76021 Karlsruhe, Germany\\
$^2$Department of Phsics, Chiba University, Chiba 263-8522, Japan\\
$^3$Institute for Low Temperature and Structure Research, 
P.O. Box 1410, 50-950 Wroc{\l}aw 2, Poland }
\date{\today}
\begin{abstract}
A theory for the Hubbard model appropriate in the limit of large $U/t$, 
small doping away from half-filling and short-ranged antiferromagnetic 
spin correlations is presented. Despite the absence of any broken symmetry
the Fermi surface takes the form of elliptical hole pockets centered near 
$(\frac{\pi}{2},\frac{\pi}{2})$ with a volume proportional to the hole
concentration. Short range antiferromagnetic correlations render the nearest 
neighbor hopping almost ineffective so that only second or third nearest 
neighbor hopping contributes appreciably to the dispersion relation. 
\end{abstract} 
\pacs{71.10.Fd,74.72.-h,71.10.Ay}
\maketitle
\section{Introduction}
The existence and shape of the Fermi surface and its change with doping
may be one of the central issues in the physics of cuprate 
superconductors.
Quantum oscillations consistent with Fermi liquid behaviour and 
a pocket-like Fermi surface have been observed in the underdoped compounds
YBa$_2$Cu$_3$O$_{6.5}$\cite{Doiron,Sebastian_1,Jaudet,Audouard}
and YBa$_2$Cu$_4$O$_8$\cite{Yelland,Bangura}.
Overdoped Tl$_2$Ba$_2$CuO$_{6+\delta}$ on the other hand
shows quantum oscillations as well
but a 'large' Fermi surface\cite{Vignolle} consistent
with band structure calculations. This indicates a change of the
Fermi surface volume around optimal doping. Similarly, 
the electron doped  compound Nd$_{2-x}$Ce$_x$CuO$_4$ shows a pocket-like
Fermi surface for electron concentrations up to $\delta=0.16$ and which changes
abruptly to a large Fermi surface for $\delta=0.17$\cite{Helm}. 
The transition in Nd$_{2-x}$Ce$_x$CuO$_4$ thus
occurs in an overdoped compound and therefore
is unrelated to antiferromagnetic ordering.\\
The validity of a Fermi liquid description in underdoped cuprates 
is incompatible with the 'Fermi arc' picture which is frequently 
invoked to describe the absence of a large Fermi surface 
in angle resolved photoemission spectroscopy (ARPES)\cite{dama}.
The Fermi surface of a Fermi liquid is a constant energy contour of the 
quasiparticle dispersion and thus necessarily a closed curve in 
${\bf k}$-space. ARPES experiments on insulating antiferromagnets like
Sr$_2$Cu$_2$O$_2$Cl$_2$\cite{Wells} and
Ca$_2$CuO$_2$Cl$_2$\cite{Ronning} have shown that the valence band
is consistent with next-nearest hopping - as in an antiferromagnet -
with maximum close to $(\frac{\pi}{2},\frac{\pi}{2})$
and that the part of the quasiparticle band facing 
$(\pi,\pi)$ has very small spectral weight. Assuming that the
effect of doping mainly consists in the chemical potential
cutting into this quasiparticle band the Fermi surface would
take the form of elliptical hole pockets and the 'Fermi arcs'
would simply be the portions of the pocket with large spectral weight. 
This has in fact been confirmed by the recent ARPES experiments
on underdoped Bi$_2$(Sr$_{2-x}$La$_x$)CuO$_6$ by Meng {\em et al.}\cite{Meng}.\\
As for the quantum oscillation experiments the electron-like nature of the 
carriers suggested by the sign of both Hall constant\cite{LeBoeuf}
and thermopower\cite{Chang} is incompatible with a straightforward 
interpretation 
in terms of hole pockets. Rather, the strong temperature dependence
of the Hall constant\cite{Rourke} and the enhancement of
low frequency magnetic excitations in a magnetic field\cite{Haug} suggest
a complicated and as yet not understood
reconstruction process to take place in
YBa$_2$Cu$_3$O$_{6.5}$ and YBa$_2$Cu$_4$O$_8$. Concerning the
thermopower it has also been pointed out that in the cuprates
there may be no obvious
correspondence between the sign of the thermopower and the Fermi surface
geometry\cite{Tranquada}.\\
In the present manuscript we investigate
the point of view that hole pockets are a generic
property of a lightly doped Mott insulator where the bulk of
electrons continues to be localized as in the insulator and
the mobile carriers correspond to the doped holes. 
The localized electrons retain only their spin degrees of freedom
and do not contribute to the volume of the Fermi surface which 
leads to a Fermi surface with a volume proportional to the hole concentration
irrespective of any broken symmetry. In fact, no experimental evidence for
any staggered order parameter which would explain hole pockets
by backfolding of a large Fermi surface has been found so far.
Moreover this picture -
a single mobile hole interacting with spin excitations - is 
the underlying one for all successful theories of the ARPES spectra of 
insulating  compounds\cite{af2,af3,af4,af5,af6,af7,af8,af9,af10}.\\
Further motivation for the present work comes from exact diagonalization 
studies of the t-J model. These show that the Fermi surface at hole dopings
$\le 15\%$ takes the form of hole pockets\cite{poc1,poc2,poc3},
that the quasiparticles have the character of strongly renormalized
spin polarons throughout this doping range\cite{r1,r2,r3} and that the low 
energy spectrum at these doping levels can be described as a Fermi liquid of 
spin $1/2$ quasiparticles corresponding to the doped holes\cite{lan}.
A comparison of the dynamical spin and density correlation function
at low ($\delta < 15\%$)
and intermediate ($\delta=30-50 \%$) hole doping moreover
indicates that around optimal doping a phase transition takes place.
In the underdoped regime spin and density correlation function differ strongly, 
with magnon-like spin excitations and extended incoherent continua in the
density correlation function\cite{den1,den} which can be explained 
quantitatively by a calculation in the spin-polaron 
formalism\cite{beckervoijta}. At higher doping, spin and density correlation 
function become more and more similar and both approach the self-convolution 
of the single-particle Green's function, whereby deviations from the 
self-convolution form can be explained as particle-hole excitations across a 
free electron-like ('large') Fermi surface\cite{intermediate}.
This rough picture would be similar to the present
experimental situation for cuprate superconductors.\\
Here we present a theory for the underdoped phase. We study the 
2 dimensional (2D) Hubbard model
\begin{eqnarray}
H&=&H_t + H_U\nonumber \\
H_t&=&-\sum_{i,j}\sum_\sigma\; t_{ij} c_{i,\sigma}^\dagger c_{j,\sigma}^{}
\nonumber \\
H_U&=& U \sum_i n_{i,\uparrow}n_{i,\downarrow}
\end{eqnarray}
where the nearest neighbor hopping $t_{10}=1$ and
we fix $U/t_{10}=8$. The results depend only weakly
on $U/t$ as long as this is sufficiently large.
In addition we use
$t_{11}=-0.1\;t_{10}$ and
$t_{21}=-t_{11}/2$. These values of $t_{11}$ and $t_{20}$
are smaller than the generally accepted ones for
cuprate superconductors - this will be discussed below.\\
In setting up a theory we have the following picture
in mind: at half-filling - i.e. the Mott insulator -
the electrons are localized and retain only their spin degrees
of freedom. The hopping term
creates charge fluctuations i.e. holes and double
occupancies which we consider as spin-$\frac{1}{2}$ Fermions.
The excitation spectrum of these Fermions has the well-known
Hubbard gap of order $U$ which exsists irrespective of any kind of order 
or broken symmetry. In section II we set up the Hamiltonian for these charge 
fluctuations. Since we really want to study the doped system where apparently 
no broken symmetry exists we thereby consider a hypothetical insulating phase
with no long range antiferromagnetic order but short ranged antiferromagnetic
spin correlations i.e. the 'spin liquid'. The 2D Hubbard model has
no broken symmetry at finite temperature so we believe it is quite
reasonable to study the charge fluctuations in such a disordered phase.
We then assume that for low doping $\delta$
this picture remains applicable, that means the holes created by
doping have the same character as the holes created by charge fluctuations 
at half-filling. Whether this assumption is justified is
a question to be answered by experiment, but we believe that the recent
experimental results lend some support to this idea.
Section III gives a summary and conclusions.
\section{Effective Hamiltonian for charge fluctuations}
\begin{figure}
\includegraphics[width=\columnwidth]{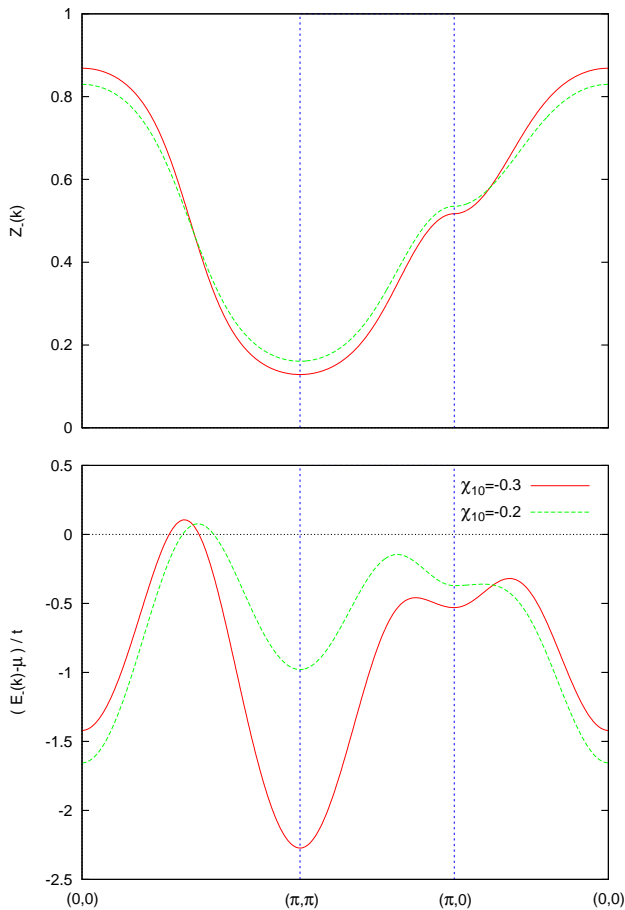}
\caption{\label{fig1}  (Color online)
Lower panel:
dispersion of the lower band at $\delta=0.1$ for different values of
$\chi_{10}$. Upper panel: dispersion of the spectral weight
of the quasiparticle band. The correlation length $\zeta=4$.}
\end{figure}

\begin{figure}
\includegraphics[width=\columnwidth]{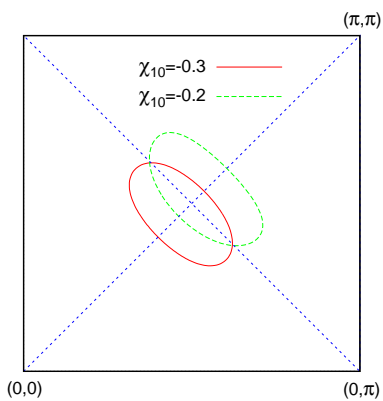}
\caption{\label{fig2}  (Color online)
Fermi surface at $\delta=0.1$ for different values of
$\chi_{10}$. The correlation length $\zeta=4$.}
\end{figure}

\begin{figure}
\includegraphics[width=\columnwidth]{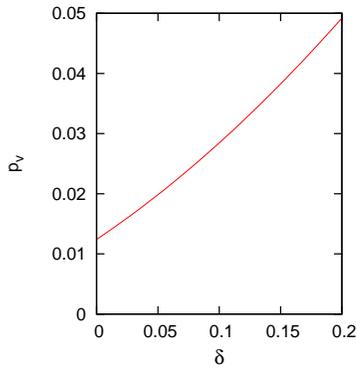}
\caption{\label{fig3}  (Color online)
Probability $p_v$ for violation of the hard-core constraint as a function
of the hole concentration $\delta$. The finite value of  $p_v$ at
$\delta=0$ is due to the charge fluctuations at half-filling.}
\end{figure}

The basic idea of the calculation is the (approximate) diagonalization of
the Hubbard Hamiltonian in a suitably chosen truncated Hilbert space.
As a starting point for constructing
the truncated Hilbert space we consider a state $|\Psi_0\rangle$ which
has exactly one electron/site, is invariant under
point group operations, has momentum zero and is a spin singlet.
These are the quantum numbers of a vacuum state and indeed
$|\Psi_0\rangle$ will play the role of a vacuum state, i.e. a state
containing neither charge nor spin fluctuations.
The only property of this state which will enter the calculation
is the static spin correlation function
\begin{equation}
\chi_{ij} = \langle \Psi_0 | {\bf S}_i \cdot {\bf S}_j |  \Psi_0 \rangle.
\end{equation}
We consider $\chi_{ij}$ as given and do not attempt to compute it.
We assume it to be antiferromagnetic and of short range i.e.
\begin{equation}
\chi_{ij}= C_0 \;e^{i {\bf Q} \cdot({\bf R}_i - {\bf R}_j)}
 e^{-\frac{|{\bf R}_i - {\bf R}_j |}{\zeta}}
\label{chi1}
\end{equation}
where ${\bf Q}=(\pi,\pi)$. Moreover $\chi_{ij}$ has to obey the constraint
\begin{equation}
\sum_{j\ne 0} \chi_{0j}= -\frac{3}{4}
\label{spincon}
\end{equation}
which follows from $|\Psi_0 \rangle$ being a singlet.
In practice we assume that (\ref{chi1}) holds only
for more distant than nearest neighbors and take
the nearest neighbor spin correlation $\chi_{10}>-0.33$
in 2D\cite{RegerYoung}
as a first free parameter. Next we choose $\zeta$ and adjust $C_0$
for the longer range part of $\chi$ so as to fulfill (\ref{spincon}).
The results for the quasiparticle dispersion turn out to be
almost independent of $\zeta$ and only weakly dependent on
$\chi_{10}$.\\
Having specified the 'spin background' $|\Psi_0 \rangle$ we introduce
the basis states of the truncated Hilbert space.
Using the familiar Hubbard operators 
$\hat{d}_{i,\sigma}^\dagger=c_{i,\sigma}^\dagger n_{i,\bar{\sigma}}^{}$
and $\hat{c}_{i,\sigma}^\dagger=c_{i,\sigma}^\dagger (1-n_{i,\bar{\sigma}}^{})$
they take the form
\begin{equation}
2^{(N_\nu + N_\mu)/2}\;
\prod_{\nu=1}^{N_\nu} \hat{d}_{i_\nu,\sigma_\nu}^\dagger\;
\prod_{\mu=1}^{N_\mu}  \hat{c}_{i_\mu,\sigma_\mu}^{}  |  \Psi_0 \rangle.
\label{basis_1}
\end{equation}
These states have double occupancies and holes
at specified positions and we treat these holes and double
occupancies as weakly interacting spin-$\frac{1}{2}$
Fermions, which is the key approximation of
the theory. Fermions are the only meanigful
description for these particles because the Hubbard operators at different
sites anticommute. Since 
$\langle \Psi_0 |\hat{d}_{i,\sigma}^{}\hat{d}_{i,\sigma}^\dagger |  \Psi_0 \rangle
= \langle \Psi_0 |\hat{c}_{i,\sigma}^\dagger\hat{c}_{i,\sigma}^{} |  \Psi_0 \rangle
=\frac{1}{2}$
the states (\ref{basis_1}) are approximately normalized
if the average distance
between the holes and double occupancies is larger
than the spin correlation length $\zeta$ (see below). This condition is 
satisfied in the limit of
large $U$, small doping and short spin correlation length,
which is the case of interest. \\
A key problem in setting up a theory for the charge fluctuations
in the Hubbard model is the peculiar nature of holes and double occupancies.
Both a hole and double occupany are spinless objects. Despite this,
for example the states $\hat{d}_{i,\uparrow}^\dagger|  \Psi_0 \rangle$ and
$\hat{d}_{i,\downarrow}^\dagger|  \Psi_0 \rangle$ are 
orthogonal. In fact, acting with $\hat{d}_{i,\sigma}^\dagger$
implies a projection onto the component of
$|\Psi_0 \rangle$ which has a $\bar{\sigma}$ electron on site $i$.
While the newly created double occupancy is a spinless object,
the information about the spin of the added electron therefore is 'stored'
in the 'spin background' and it is in a sense stored within
a spatial region of extend $\zeta$ around the respective
double occupany/hole. This can be seen by considering
expressions like
\begin{eqnarray}
\langle  \Psi_0 | \hat{c}_{i,\uparrow}^\dagger \hat{d}_{j,\uparrow}^{}  
\hat{d}_{j,\uparrow}^\dagger \hat{c}_{i,\uparrow}^{} |\Psi_0 \rangle
  &=& \frac{1}{4} - \frac{1}{3}\chi_{ij}\nonumber \\
\langle  \Psi_0 | \hat{c}_{i,\downarrow}^\dagger \hat{d}_{j,\uparrow}^{}  
\hat{d}_{j,\uparrow}^\dagger \hat{c}_{i,\downarrow}^{} |\Psi_0 \rangle
  &=& \frac{1}{4} + \frac{1}{3}\chi_{ij}\nonumber \\
\langle  \Psi_0 | \hat{c}_{i,\uparrow}^\dagger \hat{d}_{j,\downarrow}^{}  
\hat{d}_{j,\uparrow}^\dagger \hat{c}_{i,\downarrow}^{} |\Psi_0 \rangle
  &=& -\frac{2}{3}\chi_{ij} \nonumber \\
\langle  \Psi_0 | \hat{c}_{i,\downarrow}^\dagger \hat{d}_{j,\downarrow}^{}
\hat{d}_{j,\uparrow}^\dagger \hat{c}_{i,\uparrow}^{} |\Psi_0 \rangle
  &=& -\frac{2}{3}\chi_{ij}
\label{over}
\end{eqnarray}
and similar ones.
If the spin correlation function $\chi$ were zero in these
and similar expressions the states (\ref{basis_1}) would indeed be
normalized. In the following
we neglect the corrections due to a finite $\chi$ in overlap matrix elements
such as (\ref{over}). This is probably the most problematic approximation
in the present theory and induces some inaccuracies - as will be discussed 
below. Once we neglect the overlap at short distances, however, the states 
 (\ref{basis_1}) are indeed normalized. \\
Finally, the holes and double
occupancies have to obey a hard-core constraint i.e. there may be
at most one particle per site. 
In the case of small $\delta$/large $U/t$, however, the density of these 
particles is small so that it is probably a good approximation
to neglect the hard core constraint, see Appendix A for
a comparison of the present theory with linear spin wave theory
where the hard core constraint between magnons is neglected as well.\\
The procedure to be applied then is quite simple:
the states (\ref{basis_1}) are represented by states
of fictitious Fermionic spin-$\frac{1}{2}$ quasiparticles
\begin{equation}
\prod_{\nu=1}^{N_\nu} d_{i_\nu,\sigma_\nu}^\dagger\;
\prod_{\mu=1}^{N_\mu}  h_{i_\mu,\bar{\sigma}_\mu}^{\dagger}|0\rangle.
\label{basis_2}
\end{equation}
This means we have hole-like 
quasiparticles $h_{i_\nu,\sigma_\nu}^{\dagger}$
and the double occupancy-like quasiparticles $d_{i_\nu,\sigma_\nu}^\dagger$.
All operators in the quasiparticle Hilbert space are defined by demanding
that they their matrix elements between
the states (\ref{basis_2}) are identical to those of the physical operators 
between the corresponding states (\ref{basis_1}). \\
We now use this procedure to set up the quasiparticle Hamiltonian.
One has
\begin{eqnarray}
\langle \Psi_0 | \hat{c}_{j,\sigma}^\dagger \hat{d}_{i,\sigma}^{} \;H_t\;|\Psi_0 \rangle 
&=& -t_{ij}(\frac{1}{4}-\chi_{ij})\nonumber \\
\langle \Psi_0 | \hat{d}_{i,\sigma}^{} \;H_t\;\hat{d}_{j,\sigma}^\dagger|\Psi_0 \rangle 
&=& -t_{ij}(\frac{1}{4}+\chi_{ij})\nonumber \\
\langle \Psi_0 | \hat{c}_{j,\sigma}^\dagger\;H_t\;\hat{c}_{i,\sigma}|\Psi_0 \rangle 
&=&  \;\;\;t_{ij}(\frac{1}{4}+\chi_{ij})
\label{raw}
\end{eqnarray}
These matrix elements describe the pair creation of a hole/double occupancy,
the propagation of a double occupancy and the propagation of a hole.
Thereby modifications due to nearby additional particles are neglected
but again this will be reasonable for small particle density and
short spin correlation length.
Denoting
\begin{eqnarray}
V_{ij} &=& -t_{ij}(\frac{1}{2}-2\chi_{ij})\nonumber \\
\tilde{t}_{ij} &=& \;\;t_{ij}(\frac{1}{2}+2\chi_{ij})
\label{fine}
\end{eqnarray}
the Hamiltonian governing the quasiparticles therefore is
\begin{eqnarray}
H&=& -\sum_{ij,\sigma}\tilde{t}_{ij}\left( d_{i,\sigma}^\dagger d_{j,\sigma}^{} - h_{i,\sigma}^\dagger h_{j,\sigma}^{}\right)\nonumber \\
&&\;\;\;+ \sum_{ij,\sigma} V_{ij}\left( d_{i,\sigma}^\dagger h_{j\bar{\sigma}}^\dagger + h_{j\bar{\sigma}}^{} d_{i,\sigma}^{}\right)\nonumber \\
&&\;\;\;\;\;\;+ U\;\sum_{i,\sigma} d_{i,\sigma}^\dagger d_{i,\sigma}^{}
\label{hamiltonian}
\end{eqnarray}
The last term takes into account the fact that each double 
occupancy increases the
energy by $U$.The extra factor of $2$ in (\ref{fine}) as compared to (\ref{raw})
takes into account the factor of $2^{(N_\nu + N_\mu)/2}$ in (\ref{basis_1}).\\
The Hamiltonian (\ref{hamiltonian}) is solved by the transformation
\begin{eqnarray}
\gamma_{{\bf k},+, \sigma}^\dagger &=&\;\; u_{{\bf k}}\;d_{{\bf k},\sigma}^\dagger
+  v_{{\bf k}}\;h_{-{\bf k}\bar{\sigma}}^{}\nonumber \\
\gamma_{{\bf k},-,\sigma}^\dagger &=&-v_{{\bf k}}\;d_{{\bf k},\sigma}^\dagger
+  u_{{\bf k}}\;h_{-{\bf k}\bar{\sigma}}^{}
\label{bogo}
\end{eqnarray}
and we obtain the energies
\begin{eqnarray}
E_\pm({\bf k}) &=& \tilde{\epsilon}_{\bf k} + \frac{U}{2}\pm W_{\bf k}
\nonumber \\
 W_{\bf k} &=& \sqrt{ ( \frac{U}{2})^2 + V_{\bf k}^2}
\label{qpes}
\end{eqnarray}
and the coefficients
\begin{eqnarray}
u_{{\bf k}} &=&\sqrt{\frac{W_{{\bf k}}+\frac{U}{2}}{2 W_{{\bf k}}}} 
\nonumber \\
v_{{\bf k}} &=&\sqrt{\frac{W_{{\bf k}}-\frac{U}{2}}{2 W_{{\bf k}}}}
\;sign(V_{\bf k})
\end{eqnarray}
Since $c_{{\bf k},\sigma}=\hat{d}_{{\bf k},\sigma} +\hat{c}_{{\bf k},\sigma}$
the representation of the electron annihilation operator in the
quasiparticle Hilbert space becomes
\begin{equation}
c_{{\bf k},\sigma} \rightarrow \frac{1}{\sqrt{2}}(d_{{\bf k},\sigma}^{} 
+ h_{-{\bf k},\bar{\sigma}}^\dagger)
\label{cdef}
\end{equation}
where the factor of $1/\sqrt{2}$ again is due to the prefactor in 
(\ref{basis_1}).
The spectral weight of the two bands therefore is
\begin{eqnarray}
Z_{\pm}({\bf k}) &=&\frac{1}{2}(u_{{\bf k}} \pm v_{{\bf k}})^2\nonumber\\
 &=&\frac{1}{2}\left( 1\pm \frac{V_{{\bf k}} }{W_{{\bf k}}}\right)
\label{weights}
\end{eqnarray}
If we set $\chi_{ij}=0$ both $\tilde{\epsilon}_{\bf k}$ and $-V_{\bf k}$
reduce to $\epsilon_{\bf k}/2$, with $\epsilon_{\bf k}$ the noninteracting
band energy and we obtain
\begin{eqnarray}
E_\pm({\bf k}) &=& \frac{1}{2}\left(\epsilon_{\bf k} + U \pm 
\sqrt{ \epsilon_{\bf k}^2 + U^2}\right),
\nonumber \\
Z_{\pm}({\bf k}) 
 &=&\frac{1}{2}\left( 1 \mp\frac{\epsilon_{\bf k}  }
{\sqrt{ \epsilon_{\bf k}^2 + U^2}}\right),
\label{hub}
\end{eqnarray}
which is identical to the result of the Hubbard-I approximation\cite{Hubbard}
at half-filling.
The present theory thus may be viewed as an extension of the Hubbard-I 
approximation to take into account the effect of finite spin correlations.
The spin correlations, however, do have a drastic effect:
for $\chi_{10}=-0.25$  the nearest neighbor hopping matrix element
$\tilde{t}_{10}$ is zero. Here one has to bear in mind that
the value of $\chi_{10}$ in the ground state of the Heisenberg
antiferromagnet on the 2D square lattice with nn exchange only
is $\approx-0.33$\cite{RegerYoung} - a slight reduction of the spin 
correlations
due to hole doping and longer range exchange 
may well produce a value very close to $-0.25$. For small
$\tilde{t}_{10}$, however, the quasiparticle dispersion is dominated
by next-nearest neighbor hopping and the quasiparticle
dispersion is 'almost antiferromagnetic'
and thus quite different from the Hubbard-I approximation.\\
The second major difference between the present theory and the
Hubbard-I and similar approximations is the way in which electrons are counted.
In the quasiparticle Hilbert space
the operator of electron number obviously is
\begin{equation}
N_e = N + \sum_{i,\sigma} ( d_{i,\sigma}^\dagger d_{i,\sigma}^{} - h_{i,\sigma}^\dagger h_{i,\sigma}^{})
\label{elnum}
\end{equation}
because the 'spin background' $|\Psi_0\rangle$ contributes
$N$ electrons (with $N$ the number of sites),
each double occupancy increases the number of electrons by one and
each hole decreases the number of electrons by one.
When applied to a quasiparticle
state of the type (\ref{basis_2}) the operator
(\ref{elnum}) therefore gives the same
electron number as the physical electron operator applied
to the corresponding Hubbard model state (\ref{basis_1}) and this
is the prescription how operators in the quasiparticle 
Hilbert space are to be constructed.
After transformation to the $\gamma$'s and treating these as
noninteracting Fermions we obtain
\begin{equation}
N_e = \sum_{{\bf k},\sigma}\;(\;
\gamma_{{\bf k},+,\sigma}^\dagger \gamma_{{\bf k},+,\sigma}^{} + 
\gamma_{{\bf k},-,\sigma}^\dagger \gamma_{{\bf k},-,\sigma}^{}\;)
- N.
\label{elnum_1}
\end{equation}
At half-filling, $N_e=N$, the lower of the two bands is
completely filled, the upper one completely empty, which
agrees with the Hubbard-I approximation.
As the systems is doped away from half-filling, however,
the Fermi surface volume has a volume which is strictly
proportional to the number of doped holes 
- i.e. one has hole pockets with a total volume of
$\delta/2$. These do not occur as a consequence of backfolding
the Brillouin zone due to any kind of broken symmetry.\\
Pockets with a volume of $\delta/2$ are different from
the Hubbard-I approximation and the
reason is that there one uses a different way to count electrons,
namely the integrated photoemission weight.
Treating the $d$ and $h$ as ordinary Fermion operators, and
using (\ref{cdef}) we obtain the integrated spectral weight as
\begin{eqnarray}
\tilde{N}_e &=& N + \frac{1}{2} 
\sum_{{\bf k},\sigma} \langle \;d_{{\bf k},\sigma}^\dagger d_{{\bf k},\sigma}^{} - h_{{\bf k},\sigma}^\dagger h_{{\bf k},\sigma}^{}\;\rangle \nonumber \\
 &&\;\;\;+ \frac{1}{2} \sum_{{\bf k},\sigma} \langle \;d_{{\bf k},\sigma}^\dagger
h_{{-\bf k}\bar{\sigma}}^\dagger + h_{-{\bf k}\bar{\sigma}}^{} d_{{\bf k},\sigma}^{}.
\;\rangle
\label{specweight}
\end{eqnarray}
which differs from (\ref{elnum}). There are several reasons for this
discrepancy: first of all we have
\begin{equation}
 \sum_{{\bf k}} \langle \;d_{{\bf k},\sigma}^\dagger h_{{-\bf k}\bar{\sigma}}^{\dagger}
\;\rangle = \sum_{i} \langle \;d_{i,\sigma}^\dagger 
h_{i\bar{\sigma}}^{\dagger} \;\rangle 
\end{equation}
The latter expectation value, however, should be zero, because a hole and a 
double occupancy cannot occupy the same site.
Even with this term omitted, however, there is an extra factor
of $1/2$ and reason is a more fundamental one,
namely the restriction of the Hilbert space.
In fact it is easy to see that the spectral weight sum-rule cannot be
applied to an approximation like the present one where the spectral weight 
is artificially concentrated in a single band.
Let us consider a Fermi liquid with quasiparticle weight $Z<1$ and
assume that one electron is removed at the Fermi energy. This implies that one 
momentum/spin crosses from the occupied to the unoccupied
side of the Fermi energy which decreases the integrated photoemission weight
by $Z$. The remaining weight of
$1-Z$ therefore must disappear from the high-energy part of
the photoemission spectrum and reappear in the high-energy part
of the inverse photoemission spectrum.  This is in fact exactly
what is seen in exact diagonalization studies of the t-J 
model: there on has a quasiparticle band
of width $\approx 2J$ and quasiparticle weight $Z\approx0.1-0.5$ while most
of the spectral weight resides in extended incoherent 
continua\cite{szc,dago1}. Upon doping the quasiparticle peak
at the top of the band crosses the chemical potential while
simultaneously spectral weight is removed from the incoherent
part of the spectrum at energies of order $t$ below the Fermi
energy. This weight reappears - in the form of multi-magnon excitations -
at energies of order $J$ above the chemical potential and 
near $(\pi,\pi)$\cite{r1}(another way is
realized in SDW mean-field theory for the 
Hubbard model, where the weight $Z$ crosses at ${\bf k}$ and the 
remaining weight $1-Z$ at ${\bf k}+{\bf Q}$).
In a theory like the present one where all spectral weight is
concentrated in one quasiparticle band, however, these
high-energy parts do not exist and it is therefore
impossible to maintain the spectral weight sum rule. In fact, for
each electron removed from the system, $Z^{-1}$ momenta would have
to cross from photoemission to inverse photoemission to maintain the sum-rule 
and it therefore is misleading to use the integrated photoemisssion
weight to determine the Fermi surface volume. We have to accept that
the spectral weight sum rule will not be fulfilled as a consequence of 
the restriction of the Hilbert space and the lack of the
incoherent part of the spectra (this may be remedied at least partially by
including the coupling to spin excitations so as to reproduce the
incoherent continua). 
Instead, the correct expression for the particle number
is given by (\ref{elnum}). \\
In many calculations
where the Hubbard-I approximation - or any
other approximation involving Hubbard operators - is applied to the
doped case, the analogue of (\ref{specweight}) is used
to calculate the electron number. This leads to the 
peculiar 'fractional' dependence of the Fermi surface volume on hole doping
because of the fractional number $Z^{-1}$ of momenta needed
to fulfill the spectral weight sum-rule.\\
We proceed with a qualitative discussion of the band dispersion.
Since we expect the theory to be valid only for
$t\ll U$ we thereby expand in terms of $t/U$ and
keep only the lowest nonvanishing order. Moreover we note that 
there are two additional small parameters. 
For the nearest neighbor spin correlation function
$\chi_{10}$ close to $-\frac{1}{4}$, the effective nearest
neighbor hopping $\tilde{t_{10}}$ is small. 
The value of $\chi_{10}$ in the GS of the 2D Heisenberg antiferromagnet
with nearest-neighbor exchange is $-0.33$\cite{RegerYoung}, 
so that a slight reduction due to doping/longer range hopping
may well produce a value which is very close to  $-\frac{1}{4}$.
Moreover,
in the CuO$_2$ plane the $(1,1)$ and $(2,0)$ hopping integrals fulfill
$t_{11}/t_{20}\approx -2$. This equation holds
if the underlying mechanism for these matrix elements
is hopping of a Zhang-Rice singlet via Cu 4s orbitals.
If the spin correlations $\chi_{11}$ and  $\chi_{20}$ do not
differ strongly we expect both
$\tilde{t}_{11}+ 2\tilde{t}_{20}=\tau$ and
$\tilde{V}_{11}+ 2\tilde{V}_{20}=\tilde{\tau}$ to be small.\\
Expanding to lowest order in $t/U$ we have
\begin{eqnarray}
E_-({\bf k}) &=& \tilde{\epsilon}_{\bf k}- \frac{V_{\bf k}^2}{U}
\end{eqnarray}
and if $\tilde{t}_{10}$ is sufficiently small we expect the
maximum of the lower band to be near $(\frac{\pi}{2},\frac{\pi}{2})$.
At half filling the GS energy is
\begin{eqnarray}
E_0&=& 2 \sum_{\bf k} (E_-({\bf k}) -\tilde{\epsilon}_{\bf k})\nonumber\\
   &=& -2zN\sum_\alpha \;
J_{\alpha}\left(\chi_{\alpha}-\frac{1}{4}\right)^2
\end{eqnarray}
Here $\alpha \in \{10, 11, 20\dots\}$ labels the different shells 
of neighbors of a given site and
$J_{\alpha}=4t_{\alpha}^2/U$.
The true expectation value of the Heisenberg antiferromagnet in the
state $|\Psi_0\rangle$ would be obtained by replacing
$-4\left(\chi_{\alpha}-\frac{1}{4}\right)^2 \rightarrow
\chi_{\alpha}-\frac{1}{4}$. For nearest neighbor hopping only
and $\chi_{1}\approx -0.33$ the above result therefore is
a factor of $2.3$ too large.
The discrepancy is due to the neglect of overlap integrals such as
(\ref{over}), see Appendix B. The inclusion of such overlap integrals
probably is only a technical problem but we have to bear in mind that the
simplification of neglecting such overlap integrals results
in inaccuracies.\\
Next we consider the quasiparticle dispersion relation. As already mentioned
if $\tilde{t}_{10}$ is small we expect the maximum of the dispersion
of the lower band near $(\frac{\pi}{2},\frac{\pi}{2})$. 
Setting $k_\alpha=\frac{\pi}{2} + \kappa_\alpha$ and expanding to
second order in  $\kappa$ the constant energy contours
are elliptical
\begin{equation}
E_-({\bf \kappa}) = E_{max} -
\frac{(\kappa_+ - \kappa_0)^2}{a^2} - \frac{\kappa_-^2}{b^2}
\end{equation}
with $\kappa_\pm = \frac{1}{\sqrt{2}}(\kappa_x \pm \kappa_y)$ and -
neglecting terms $\propto \tau, \tilde{\tau}$ as well as
terms of higher order in $t/U$ - one finds
\begin{eqnarray}
\frac{1}{a^2} &=&\;\; 2 J_{10}\;(\frac{1}{2}-2\chi_{10})^2\nonumber \\
\frac{1}{b^2} &=&-4 t_{11}(\frac{1}{2}+2\chi_{11})
\nonumber\\
\frac{\kappa_0}{a^2} &=& \sqrt{2} \left( t_{10}\;(\frac{1}{2}+2\chi_{10})
+J_{10}\frac{t_{11}}{t_{10}}(\frac{1}{2}-2\chi_{10})
(\frac{1}{2}-2\chi_{11})\right)\nonumber\\
\end{eqnarray}
Depending on the sign of
$\chi_{10}+\frac{1}{4}$ the pockets thus are shifted towards
$\Gamma$ or away from $\Gamma$.\\
It turns out that the correlation length $\zeta$ has negligible
influence on the dispersion. More significant is the value
of the nearest neighbor spin correlation function $\chi_{10}$.
Figure \ref{fig1} shows the quasiparticle dispersion and the
spectral weight of the band for two values of $\chi_{10}$.
Figure \ref{fig2} shows the pockets obtained for the two different values of
$chi_{10}$, which obviously determines the position of the hole pocket in 
the Brillouin zone. The pockets are not centered
at $(\frac{\pi}{2},\frac{\pi}{2})$,
for the larger values of $\chi_{10}$ the pocket is shifted towards
$(0,0)$  as it is seen in the ARPES experiment\cite{Meng}. 
The overall shape of the dispersion relation is quite similar as the
one for a hole in an 
antiferromagnet\cite{Bula,Trugman,Shraiman,Inoue,Ederbecker} 
but it should be noted that
the present theory does not use any antiferromagnetic order.
The 'antiferromagnetic shape' of the dispersion is due to the fact
that even short range
antiferromagnetic correlations combined with the strong Coulomb
repulsion between electrons are sufficient to suppress the nearest neighbor
hopping almost completely. More detailed
calculations show that antiferromagnetic spin
correlations instead enhance the incoherent
nearest neighbor hopping, i.e. nearest neighbor hopping
involves emisssion or absorption of a spin excitation\cite{tobepub}.
It also should be noted that the spectral weight of the 'backside'
of the pocket, i.e. the part facing $(\pi,\pi)$ is low.
This is also seen in ARPES\cite{Wells,Ronning,Meng}, 
although the difference of spectral weight is much more pronounced there.
By analogy with hole motion in an antiferromagnet
one may assume, however, that the coupling to spin excitation and
formation of spin polarons will enhance the
difference in spectral weight and thus make the theory more similar
to experiment\cite{wrobel}.\\
It should be noted that the antiferromagnetic correlations
determine the location of the pocket in the Brillouin zone
but not the volume of the Fermi surface, which is given
by (\ref{elnum_1}). As already mentioned, if one sets the spin correlation
function $\chi=0$ - which actually violates the singlet condition
(\ref{spincon}) - the dispersion relation agrees with that of the
Hubbard I approximation and the hole pocket is then centered
at $(\pi,\pi)$. A hole pocket around $(\pi,\pi)$ has indeed
been observed in Quantum Monte Carlo (QMC) simulations of the Hubbard
model\cite{Carsten} and it is plausible that the high temperature used in the
(QMC) simulation renders the spin correlation function small or zero and thus
shifts the pockets to the corner of the Brillouin zone.\\
Finally we note that 
the present theory tends to overestimate the impact of the
$t_{11}$- and $t_{20}$-terms on the quasiparticle dispersion
which is why we used relatively small
values of  $t_{11}$ and $t_{20}$ used here. This is probably related
the fact that no coupling to spin excitations is taken into
account in the present theory which increases the quasiparticle weight and the
effect of the $t_{11}$- and $t_{20}$-terms.\\
To conclude this section we return to the issue of
the hard-core constraint between the holes/double occupancies.
To that end we consider the total densities of 
holes/double occupancies per spin direction:
\begin{eqnarray}
n_d&=&\frac{1}{N}\;\sum_{i} \langle\; d_{i,\sigma}^\dagger d_{i,\sigma}^{} \; \rangle\nonumber\\
   &=&\frac{1}{N}\; \sum_{\bf k} v_{\bf k}^2 \;f(E_-({\bf k}))\nonumber\\
n_h&=&\frac{1}{N}\;\sum_{i} \langle\;h_{i,\sigma}^\dagger h_{i,\sigma}^{} \; \rangle\nonumber\\
   &=& \frac{1}{N}\;\sum_{\bf k}\left( v_{\bf k}^2 + u_{\bf k}^2 
(1- f(E_-({\bf k})))\right)\nonumber\\
\end{eqnarray}
This may serve as a criterion for the quality of the approximation
to relax the hard-core constraint. Namely the probability for violation
of the constraint at a given site is
\begin{equation}
p_v=4 n_d n_h + n_d^2 + n_h^2
\end{equation}
and this is shown in Figure \ref{fig3} for 
$U/t=8$ as a function of the hole concentration $\delta$.
This implies that even at $\delta=0.2$ the constraint
is violated at $5\%$ of the sites. Enforcement of the
constraint e.g. by Gutzwiller projection would therefore have a small
influence on the results. The neglect of the constraint therefore is
probably a quite reasonable approximation.\\
\section{Summary and Discussion}
In summary a theory for the lightly doped Mott insulator has been
derived. The basic assumption thereby is that holes
introduced by doping have the same nature
as the hole-like charge fluctuations at half-filling
whose density is $\propto (\frac{t^2}{U^2})$. For low hole doping this
is probably a reasonable assumption. The quasiholes then form a Fermi gas 
with a total Fermi surface volume of $\delta/2$.
Antiferromagnetic spin correlations render the nearest neighbor hopping 
essentially ineffective so that the
dispersion relation is similar to the one for hole motion in an antiferromagnet
even in the complete absence of static antiferromagnetic order.
The Fermi surface thus takes the form of four elliptical hole pockets
centered near $(\frac{\pi}{2},\frac{\pi}{2})$. In addition to these
Fermionic excitations the doped insulator probably has a second type
of excitations, namely Bosonic spin triplet excitations 
which are similar in character as the magnons at half-filling.
We postpone the discussion of these excitations and their
interaction with the charge fluctuations\cite{tobepub}.\\
A number of simplifying assumptions of different quality were made: the spin 
correlation function of the 'spin background' was assumed to have a simple 
form and was taken as a given input parameter. It would be desirable to 
calculate this e.g. be minimization of the total energy but this would
require a solution of the system of interacting quasiparticles
and magnons. On the other hand the form of the spin correlation function
which was assumed seems quite physical and moreover the
results do not change strongly with the parameters of the
spin correlation funtion. For example the quasiparticle dispersion is 
essentially independent of the spin correlation legth $\zeta$.\\
In solving the Hamiltonian for the charge fluctuations the overlap
between pairs of particles has been neglected. This is probably the most
drastic approximation made and was shown to lead to inaccuracies in the 
results e.g. a deviation from the ground state energy at half-filling from 
the known energy of the Heisenberg antiferromagnet. The neglected overlaps - 
being 'four particle overlaps' - would create an interaction between the 
quasiparticles.
The neglect of the hard-core constraint between the particles, on the other 
hand, is probably a very reasonable  approximation because the density 
of the charge fluctuations is small.\\
An interesting question and possibly the key to understand high-temperature
superconductivity is the nature of the phase transition between this 
correlation-dominated low doping 
phase with a hole-pocket-like Fermi surface
and the intermediate and electron-density phase with a 'large' Fermi surface
which has been inferred e.g. from the dynamical spin and density correlation 
function\cite{intermediate}. Experimental data suggest
that this phase transition occurs at optimal doping or
in the overdoped range of hole concentrations and thus
is related to the mechanism of superconductivity. In the framework
of the present formalism this might
correspond to a replacement of a spin-liquid
like 'spin-background' $|\Psi_0\rangle$
to e.g. a Gutzwiller-projected Fermi sea
where the spin correlation function has long-ranged Friedel-like
oscillations. These would introduce long-ranged overlap integrals
between the quasiparticles and thus enhance their interaction.\\
Acknowledgement: R. E. most gratefully acknowledges the kind hospitality
at the Center for Frontier Science, Chiba University.
\section{Appendix A}
In this Appendix a re-derivation of linear spin wave for the
spin-$\frac{1}{2}$ Heisenberg antiferromagnet is given to show the
analogy with the present theory for the Hubbard model and to some extent
justify the neglect of the hard-core constraint. For spin wave theory
the role of
$|\Psi_0\rangle$ is played by the N\'eel state, $|\Psi_N\rangle$,
and the misalligned spins or magnons play the same role as
the charge fluctuations in the Hubbard model.
The misalligned spins are represented by Boson operators $a_i^\dagger$
and $b_i^\dagger$ which are defined on the $\uparrow$- and 
$\downarrow$-sublattices, respectively.
The $a_i^\dagger$ and $b_i^\dagger$ must be chosen as Bosons because
the operations of inverting spins at different sites commute.
These anticommutation relations do not hold
for operators referring to the same site $i$ - rather, for the
spin-$\frac{1}{2}$ system the $a^\dagger$ and $b^\dagger$ Bosons have to obey
a hard-core constraint because a spin can be flipped only once.
Each misalligned spin increases the energy
by $\frac{zJ}{2}$ whence we have the representation of the longitudinal part:
\[
H_0= \frac{zJ}{2}
\left(\sum_{i\in A} a_{i}^\dagger a_i^{} + \sum_{j\in B} b_{j}^\dagger b_j^{}\right)
\]
where $z$ is the number of nearest neighbors and $J$ the exchange constant.
The transverse part of the Heisenberg exchange
creates or annihilates pairs of spin fluctuations:
\[
H_1= \frac{J}{2}\sum_{\langle i,j\rangle} \left(a_{i}^\dagger b_{j}^\dagger + 
H.c.\right).
\]
where the sum is over all pairs of nearest neighbors.
Adding the two terms gives the familiar spin-wave Hamiltonian with
 $a^\dagger$ and $b^\dagger$ still having to obey a hard-core-constraint.
This derivation is completely analogous as for the
charge fluctuations in the Hubbard model. In linear spin wave theory
the hard-core constraint between the Bosons
is now simply ignored and the $a^\dagger$ and $b^\dagger$ operators
are treated as free Boson operators. Despite this, linear spin wave
theory is a highly successful theory and the reason is
that the density $n$ of Bosons in the ground state - obtained 
self-consistently from the solution of the
spin wave Hamiltonian itself - is low. Even for the 2D Heisenberg
antiferromagnet one has $n=0.197$ so that the probability
that two Bosons occupy the same site and thus violate the hard core
constraint is only $n^2\approx 4\%$. Relaxing the constraint thus
will be a very good approximation. In the limit of large $U$ and low
doping the density of charge fluctuations will be small as well
(see Figure \ref{fig3})
and we expect that relaxing the hard-core constraint for the Fermions
will be a reasonable approximation as well.
\section{Appendix B}
In this Appendix we show that by properly taking into account
the overlap integrals the correct expectation value
of the energy of the Heisenberg antiferromagnet can be obtained.
We consider half-filling and start with the
state $|\Psi_0\rangle$. We choose two sites,
$i$ and $j$ which are connected by
the hopping term. By acting with the pair creation part $\propto V_{ij}$
for this bond term we can generate the states
$|1\rangle=d_{i,\uparrow}^\dagger h_{j,\downarrow}^\dagger|0\rangle$ and
$|2\rangle=d_{i,\downarrow}^\dagger h_{j,\uparrow}^\dagger|0\rangle$
More precisely, the pair creation part generates the state
$|1\rangle + |2\rangle$. Using the overlap integrals
in (\ref{over}) it is straightforward to see that this state
is an eigenstate of the overlap matrix $N_{ij}=\langle i|j\rangle$
with eigenvalue $(1-4\chi_{ij})$ (there is a factor of $4$ due to the
prefactor in (\ref{basis_1})) and therefore
has the norm $n=2(1-4\chi_{ij})$. The matrix element between $|\Psi_0\rangle$
and the normalized state $(1/\sqrt{n})(|1\rangle + |2\rangle)$ then is
\[
-t_{ij}\sqrt{\frac{1-4\chi_{ij}}{2}}
\]
so that 2$^{nd}$ order pertubation theory gives the energy
per bond $J_{ij}(\chi_{ij}-\frac{1}{4})$. Here the additional
factor of $2$ comes from the analogous process where the double
occupancy is created at $j$ and the hole at $i$.

\end{document}